\documentclass[aps,prl,superscriptaddress,showpacs,floatfix,twocolumn]{revtex4-1}
\usepackage{times}
\usepackage{graphicx,graphics,color}
\usepackage{amsmath}
\usepackage{amssymb}
\usepackage{ulem}

\begin{document}

\title{A unified description of superconducting pairing symmetry in electron-doped Fe-based-122 compounds}

\author{Bo Li}
\affiliation{Department of Physics and Texas Center for Superconductivity, University of Houston, Houston, Texas 77204, USA}

\author{Lihua Pan}
\affiliation{Department of Physics and Texas Center for Superconductivity, University of Houston, Houston, Texas 77204, USA}
\affiliation{School of Physics Science and Technology, Yangzhou University, Yangzhou 225002, China}

\author{Yuan-Yen Tai}
\affiliation{Theoretical Division, Los Alamos National Laboratory, Los Alamos, New Mexico 87545, USA}

\author{Matthias J. Graf}
\affiliation{Theoretical Division, Los Alamos National Laboratory, Los Alamos, New Mexico 87545, USA}

\author{Jian-Xin Zhu}
\affiliation{Theoretical Division, Los Alamos National Laboratory, Los Alamos, New Mexico 87545, USA}
\affiliation{Center for Integrated Nanotechnologies, Los Alamos National Laboratory, Los Alamos, New Mexico 87545, USA}

\author{Kevin E. Bassler}
\affiliation{Department of Physics and Texas Center for Superconductivity, University of Houston, Houston, Texas 77204, USA}

\author{C. S. Ting}
\affiliation{Department of Physics and Texas Center for Superconductivity, University of Houston, Houston, Texas 77204, USA}

\date{\today}

\begin{abstract}
The pairing symmetry is examined in highly electron-doped Ba(Fe$_{1-x}$Co$_x$As)$_2$ and A$_y$Fe$_2$Se$_2$ (with A=K, Cs) compounds, with similar crystallographic and electronic band structures.
Starting from a phenomenological two-orbital model, we consider nearest-neighbor and next-nearest-neighbor intraorbital pairing interactions on the Fe square lattice.
In this model, we find a unified description of the evolution from $s_\pm$-wave pairing ($2.0 < n \lesssim 2.4$) to $d$-wave pairing ($2.4 \lesssim n \lesssim 2.5$) as a function of electron filling. In the crossover region a novel time-reversal symmetry breaking state with $s_\pm+id$ pairing symmetry emerges.
This minimal model offers an overall picture of the evolution of superconductivity with electron doping for both $s_\pm$-wave [Ba(Fe$_{1-x}$Co$_x$As)$_2$] and  $d$-wave [A$_y$Fe$_2$Se$_2$] pairing, as long as the
dopants only play the role of a charge reservoir.
However, the situation is more complicated for Ba(Fe$_{1-x}$Co$_x$As)$_2$.
A real-space study further shows that when the impurity scattering effects of Co dopants are taken into account, the superconductivity is completely suppressed for $n > 2.4$. This preempts any observation of $d$-wave pairing in this compound, in contrast to A$_y$Fe$_2$Se$_2$.
\end{abstract}


\maketitle

The recent discovery of Fe-pnictide based superconductors offers a new family of materials in which the nature of superconductivity can be explored~\cite{Kamihara,Zhi-An,XHChen,Cruz,GFChen}.
Although the mechanism of superconductivity (SC) in this family still remains an open subject, several theoretical models indicate that the complex geometry of the Fermi surface (FS), including both hole and electron pockets in the Brillouin zone (BZ), should be mainly responsible for the SC with $s_\pm$-wave pairing symmetry in weakly and moderately doped systems~\cite{Mazin,Wang,Yao,Seo,Kuroki}.
In the FeAs-$122$ family, the SC in the electron-doped Ba(Fe$_{1-x}$Co$_x$As)$_2$ disappears almost at the same doping level where the hole-FS pockets vanish near the $\Gamma$-point~\cite{Sekiba}.
This feature has been widely interpreted to indicate the existence of a correlation  between the SC and the FS topology of the Fe-pnictide compounds.
On the other hand, a new series of iron-chalcogenide $122$-compounds such as A$_y$Fe$_2$Se$_2$ (A=K, Cs), have been discovered recently with relatively high SC transition temperature ($T_c$) of about $31$ K~\cite{JGuo,JJYing,WLi,XLC}.
These superconductors are the most heavily electron-doped among the iron-based compounds with $0.8 \lesssim y \leq 1$.
Electronic band structure calculations~\cite{LZhang,XWYan,CCao,Shein,Nekrsov} for A$_y$Fe$_2$Se$_2$ indicate that only electron pockets exist near the M-point of the BZ.
A recent angle-resolved photoemission spectroscopy (ARPES) experiment~\cite{19YZhang} also showed the presence of electron pockets around the M point and the near absence of a hole pocket around the $\Gamma$-point at $y=0.8$.
For weakly and moderately electron-doped Ba(Fe$_{1-x}$Co$_x$As)$_2$ with $x<0.4$, the FS consists of both electron pockets near M points and hole pockets near the $\Gamma$ point.
These compounds are thought to have $s_\pm$-wave pairing symmetry due to magnetic fluctuations between the M and $\Gamma$ points~\cite{Mazin} on the FS.
However, for the heavily electron-doped A$_y$Fe$_2$Se$_2$ with $y=0.8$ to $1$, the FS contains no hole pockets at the $\Gamma$ point and the proper superconducting pairing symmetry should not be $s_\pm$-wave. Instead, it was predicted to have $d$-wave pairing symmetry~\cite{JXZhu,RYu,TDas,TAMaier,FWang}.
Therefore, it is very important to understand the evolution of the superconducting pairing symmetry within a real-space formulation for  A$_y$Fe$_2$Se$_2$ with $0.8 \lesssim y \leq 1$ and at the same time why SC completely disappears for Ba(Fe$_{1-x}$Co$_x$As)$_2$ with $x>0.4$.

So far the $s_\pm$-wave pairing symmetry in Ba(Fe$_{1-x}$Co$_x$As)$_2$ has been attributed to a next-nearest-neighbor (NNN) pairing interaction among the electrons of the Fe atoms~\cite{Seo,20DZhang}. 
In addition, its phase diagram has been mapped out as a function of electron doping~\cite{22TZhou,YYTai,HChen}
and it has been shown that SC  vanishes for electron doping $x>0.4$.
This interpretation seems plausible, since at this point the hole pockets at the $\Gamma$ point vanish as the hole states are completely filled by doped electrons.
Not only are these results consistent with experiments~\cite{23Pratt,24Lester,25WFWang,26Laplace}, they also suggest that the NNN-pairing interaction is able to capture the essential ingredients of the magnetic fluctuations between the M and $\Gamma$ points~\cite{Mazin}.
According to the above theoretical works~\cite{22TZhou,YYTai,HChen}, there is no SC in A$_y$Fe$_2$Se$_2$ with $0.8 \lesssim y \leq 1$, in clear contradiction with recent experiments.

In this work, we seek to obtain a unified picture of the evolution of the superconducting pairing symmetry with moderate pairing strength for the above mentioned compounds by including the nearest-neighbor (NN)-pairing interaction $V_{\text{NN}}$ in addition to the NNN-pairing interaction $V_{\text{NNN}}$.
Since the electronic structure of A$_y$Fe$_2$Se$_2$ is similar to that of Ba(Fe$_{1-x}$Co$_x$As)$_2$, we will employ this new Hamiltonian to re-examine the SC of both materials.
The strength of the pairing interactions is chosen according to two factors, one is that the pairing symmetry of Ba(Fe$_{1-x}$Co$_x$As)$_2$ is dominantly $s_\pm$-wave at doping level $x<0.4$, the other one is that the A$_y$Fe$_2$Se$_2$ has a finite SC order parameter at doping level $0.8 \lesssim y \leq 1$.
The Hamiltonian is numerically solved on a $28\times28$ lattice by using the Bogoliubov-de-Gennes (BdG) equations.
Our results reveal that the pairing symmetry changes from $s_\pm$-wave to $d$-wave as the electron doping is increased.
For moderately electron-doped Rb$_y$Fe$_2$Se$_2$ (with $y=0.3$), the experimentally observed SC~\cite{27YTexier} should have an $s_\pm$-wave pairing symmetry.
In Ba(Fe$_{1-x}$Co$_x$As)$_2$ compound, although the SC is predicted for $x>0.4$ in the present calculation, it has not been observed in experiments~\cite{23Pratt,24Lester,25WFWang,26Laplace}.
We will numerically demonstrate that the $d$-wave SC in this region can be completely suppressed by scattering due to randomly distributed Co atoms in the Fe planes.

We use a phenomenological two-orbital tight-binding model~\cite{HChen} to numerically perform calculations of the FS, local density of states (LDOS), magnetic and SC order parameters.
This model has been successfully used to theoretically describe the generic phase diagram and other properties of FeAs-$122$  superconductors~\cite{28LPan,HChen,29LPan,BLi}.
Based on the fact that they share similar crystal and band structures with the FeAs-$122$ family, it is reasonable to apply our two-orbital model, which is built to describe the FeAs-$122$ family, to A$_y$Fe$_2$Se$_2$ compounds, after an additional NN-pairing interaction is considered.

Consider the Hamiltonian $\mathcal{H}=H_0+H_{SC}+H_{int}$ that describes the energy of charge carriers.
Without the impurity term, here $\mathcal{H}$ follows the same formulation as Ref.~\onlinecite{HChen} for the hopping ($H_0$), pairing ($H_{SC}$) and on-site interaction ($H_{int}$) terms.
We express the matrix form of $\mathcal{H}$ with the basis, $\psi_{i\mu}=(c_{i\mu\uparrow}, c_{i\mu\downarrow}^\dagger)^\text{Transpose}$,
$\mathcal{H}=\sum_{i\mu j\nu} \psi_{i\mu}^\dagger \, H_{\text{BdG}} \,\psi_{j\nu}$
to calculate the eigenvalue and eigenvectors of $H_{\text{BdG}}$,
\begin{eqnarray}
\sum_{{j},\nu}\left( \begin{array}{ccc}
H_{{i}\mu{j}\nu\uparrow}& \delta_{\mu\nu}\Delta_{{i}\mu{j}\nu}
\\ \delta_{\mu\nu}\Delta^{*}_{{i}\mu{j}\nu}&-H^{*}_{{i}\mu{j}\nu\downarrow}
\end{array} \right)  \left( \begin{array}{ccc}
u^{n}_{{j}\nu\uparrow} \\ v^{n}_{{j}\nu\downarrow}
\end{array} \right)=E_n\left(
\begin{array}{ccc}u^{n}_{{i}\mu\uparrow} \\
v^{n}_{{i}\mu\downarrow} \end{array} \right)
\end{eqnarray}
here both NN and NNN intra-orbial pairing orders are calculated from the following equations~\cite{HChen},
\begin{eqnarray}
\Delta^{\epsilon}_{{i}\mu{j}\nu}=\frac{V_{\epsilon}}{4}\sum_{n}(u^{n}_{{i}\mu\uparrow}
v^{n*}_{{j}\nu\downarrow} +u^{n}_{{j}\nu\uparrow}
v^{n*}_{{i}\mu\downarrow})\tanh\frac{E_n}{2k_BT},
\end{eqnarray}
where $\epsilon\in$ \{NN, NNN\} to denote the nearest-neighbor and next-nearest-neighbor pairing bond.
Throughout this work, we use the same hopping and on-site interaction as in Ref.~\onlinecite{HChen}. The chemical potential is determined self-consistently calculating the averaged electron filling $n$ from the BdG equations.
$V_{\text{NN}}$ is the intra-orbital pairing interaction between NN sites, and $V_{\text{NNN}}$ is the intra-orbital pairing interaction between NNN sites.
All energies are measured in units of nearest-neighbor intra-orbital hopping $\left|t_5\right|$, see definition in Ref.~\onlinecite{HChen}.
The collinear spin density wave (SDW) order parameter is defined as
$m_{{i}}=(-1)^{i_x}\frac{1}{4}\sum_{\mu}(\langle n_{{i}\mu\uparrow}\rangle-\langle n_{{i}\mu\downarrow}\rangle)$,
where the $i_x$ is the lattice number along x-axis and $\langle n_{i\mu\sigma} \rangle$ is the electron density for site-$i$ orbital-$\mu$ and spin-$\sigma$.
The pairing order parameters of NN $d$-wave and NNN $s_{\pm}$-wave are defined as\\
$\bullet\;\;\Delta_{d}=\frac{1}{8N}|\sum_{ij\mu\nu}\epsilon_x\epsilon_y\Delta^{\text{NN}}_{{i}\mu{j}\nu}|$ and\\
$\bullet\;\;\Delta^{\prime}_{s_\pm}=\frac{1}{8N}\sum_{ij^{\prime}\mu\nu}\Delta^{\text{NNN}}_{{i}\mu{j^{\prime}}\nu}$, respectively,\\
where $\textbf{j}=\textbf{i}\pm\hat{x}(\hat{y})$ is the NN sites of site $\textbf{i}$ and $\textbf{j}^{\prime}=\textbf{i}\pm\hat{x}\pm\hat{y}$ is the NNN sites of site $\textbf{i}$, $\epsilon_{x(y)}=(\textbf{j}^{(\prime)}-\textbf{i})\cdot\hat{x}(\hat{y})$ and $N$ is the number of Fe lattice sites.
We can write down the correspond form factor in k-space for $\Delta_d$ and $\Delta^\prime_{s_\pm}$ under 2-Fe per unit cell Brillouin zone\\
$\bullet\;\; \Delta_{d}(k)=2\Delta_{d}[\sin(k_x)\sin(k_y)]$ and\\
$\bullet\;\; \Delta^\prime_{s_\pm}(k)=2\Delta^\prime_{s_\pm}[\cos(k_x)+\cos(k_y)]$.\\
In two of our previous works~\cite{YYTai,28LPan}, it was shown that the $s_\pm$-wave superconductivity in the phase diagram for Ba(Fe$_{1-x}$Co$_x$As)$_2$ can be constructed from the NNN intraorbital pairing interaction.
The approach in these works~\cite{YYTai,28LPan} gives a unified description only for $s_\pm$-wave pairing of the entire phase diagram covering both the electron and hole doped regimes.
The k-space $s_\pm$-wave SC order parameter vanishes in highly doped regime ($n\ge2.4$) where the hole pocket on the FS at the $\Gamma$-point shrinks to zero.
On the other hand, SC has been observed in recently discovered A$_x$Fe$_2$Se$_2$ at higher-doping level than $n=2.4$~\cite{JGuo,JJYing,WLi,XLC} with a predicted $d$-wave pairing symmetry,  has been studied under k-space band picture~\cite{JXZhu,RYu,TDas,TAMaier,FWang}.
An insight of real-space picture of the $d$-wave pairing in addition to what was used in the previous works should be taken into consideration.

\begin{figure}
 \includegraphics[scale=0.45,angle=0]{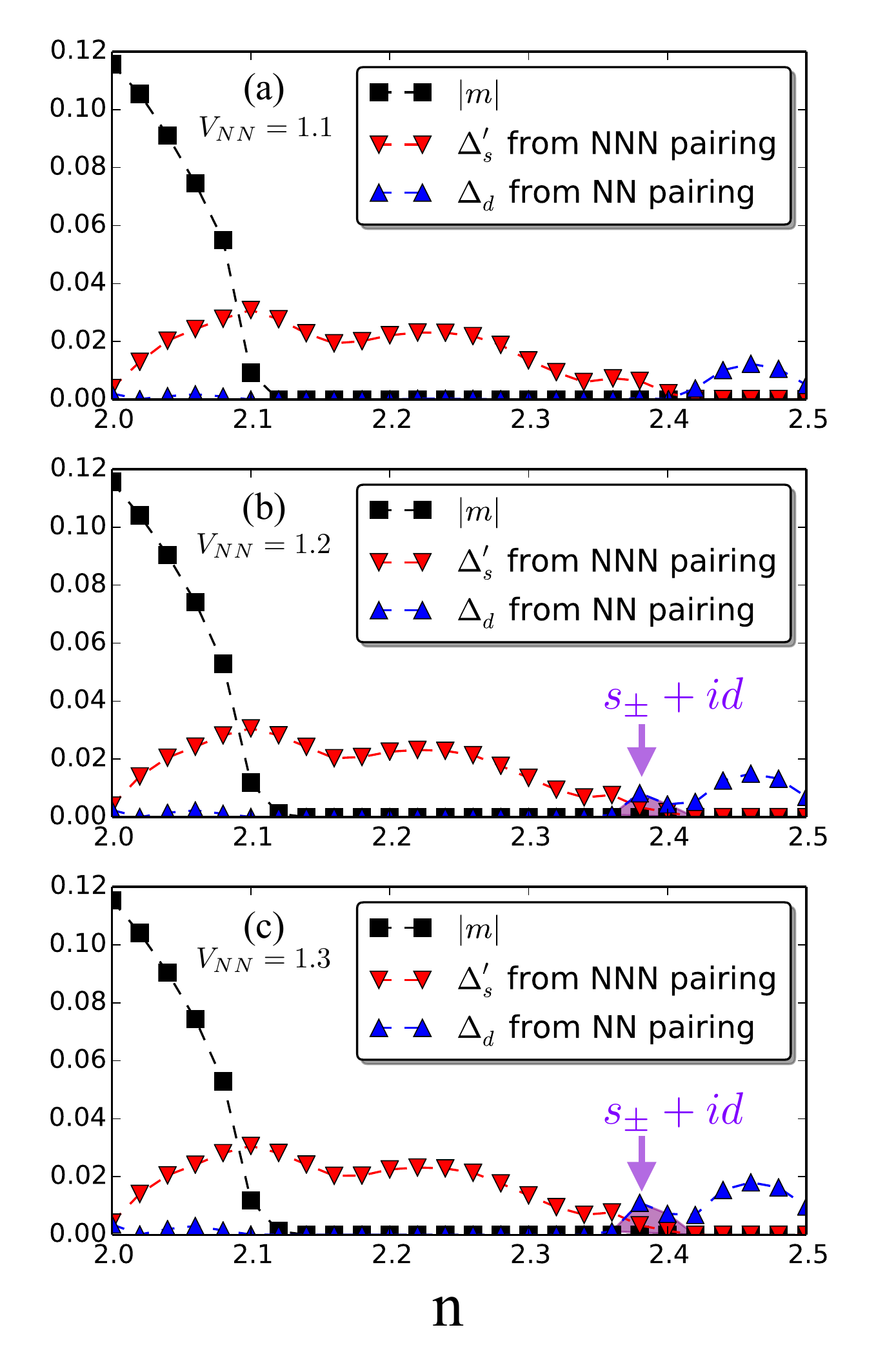}
  \caption
   {(Color online) SC and AFM phase diagram at $T=0K$ for uniformly doped A$_y$Fe$_2$Se$_2$ and  Ba(Fe$_{1-x}$Co$_x$As)$_2$ as a function of the averaged electron number per site $n$, with NN pairing $V_{\text{NN}}$ interaction chosen to be (a) $1.1$ (b) $1.2$ and (c) $1.3$. Black squares, red and blue triangles represent the collinear SDW order parameter, pairing order parameter of NNN $s_\pm$-wave and NN $d$-wave, correspondingly.
   The NNN pairing interaction is set to be fixed at $V_{\text{NNN}}=1.05$. The purple shaded region represents the time-reversal breaking pairing state, $s_\pm+id$.}
   \label{phase1}
\end{figure}

In order to seek a unified theme in the real-space lattice model for the pairing symmetry of the electron doped A$_y$Fe$_2$Se$_2$ and other FeAs-122 compounds, we find the NN intraorbital pairing interaction is the last piece of jigsaw to complete the entire picture.
The phase diagrams thus obtained are shown in Fig.~\ref{phase1} for fixed $V_{\text{NNN}}$ and several $V_{\text{NN}}$.
The average electron-number per site is defined as $n=2+x=2+y/2$.
The coexistence of SC order parameters from different pairing symmetry are labeled in the figure, together with $2\times1$ collinear SDW order.
In the calculation the NNN pairing is set to be constant $V_{\text{NNN}}=1.05$, but the NN pairing is changing from $V_{\text{NN}}=1.1$ to $1.3$.
When $V_{\text{NN}}=1.1$ as shown in Fig.~\ref{phase1}(a), from $n=2.0$ to $n=2.4$, the pairing symmetry is dominantly $s_\pm$-wave and the $d$-wave component is very weak, while in the doping region $2.4<n<2.5$, the SC has a dominatly $d$-wave like symmetry originating in the NN pairing interaction.
There exists a sharp transition of the SC pairing symmetry at $n=2.4$.
When $V_{\text{NN}}$ is increased to $1.2$ as shown in Fig.~\ref{phase1}(b), the $d$-wave dominant region expands down to $n=2.38$ where pairing symmetry becomes a mixed complex one, $s_\pm+id$ ($2.38<n<2.4$), meanwhile the magnitude the pure $d$-wave SC order increased a bit in the highly electron doped region ($2.4<n<2.5$).
The SC and AFM orders remain unchanged at all other doping levels.
When the NN pairing interaction is further increased to $1.3$ as shown in Fig.~\ref{phase1}(c), the complex $d$-wave SC order further expands to lower doping at $n=2.36$, but still enhanced in the highly electron-doped region.
The magnetic order and the SC pairing symmetry in Fig.~\ref{phase1}(c) at other doping levels remain the same as the previous two figures.
In both the $V_{\text{NN}}=1.2$ and $1.3$ cases, the regions for pairing symmetry transitions become broadened and the pairing symmetry becomes a mixed complex one, $s_\pm+id$, for $2.36\lesssim n<2.4$. If the NN pairing interaction is larger than $1.3$, the $d$-wave pairing order will emerge in optimal-doped region or even in the under-doped region. This is clearly not in agreement with experiments performed on Ba(Fe$_{1-x}$Co$_x$As)$_2$.
It is also not reasonable to choose $V_{\text{NN}}<V_{\text{NNN}}=1.05$, because the obtained $d$-wave SC order parameter in A$_y$Fe$_2$Se$_2$ (with $y\sim0.8$ to $1$) would be too small to explain the experiments.

\begin{figure}
 \includegraphics[scale=0.4,angle=0]{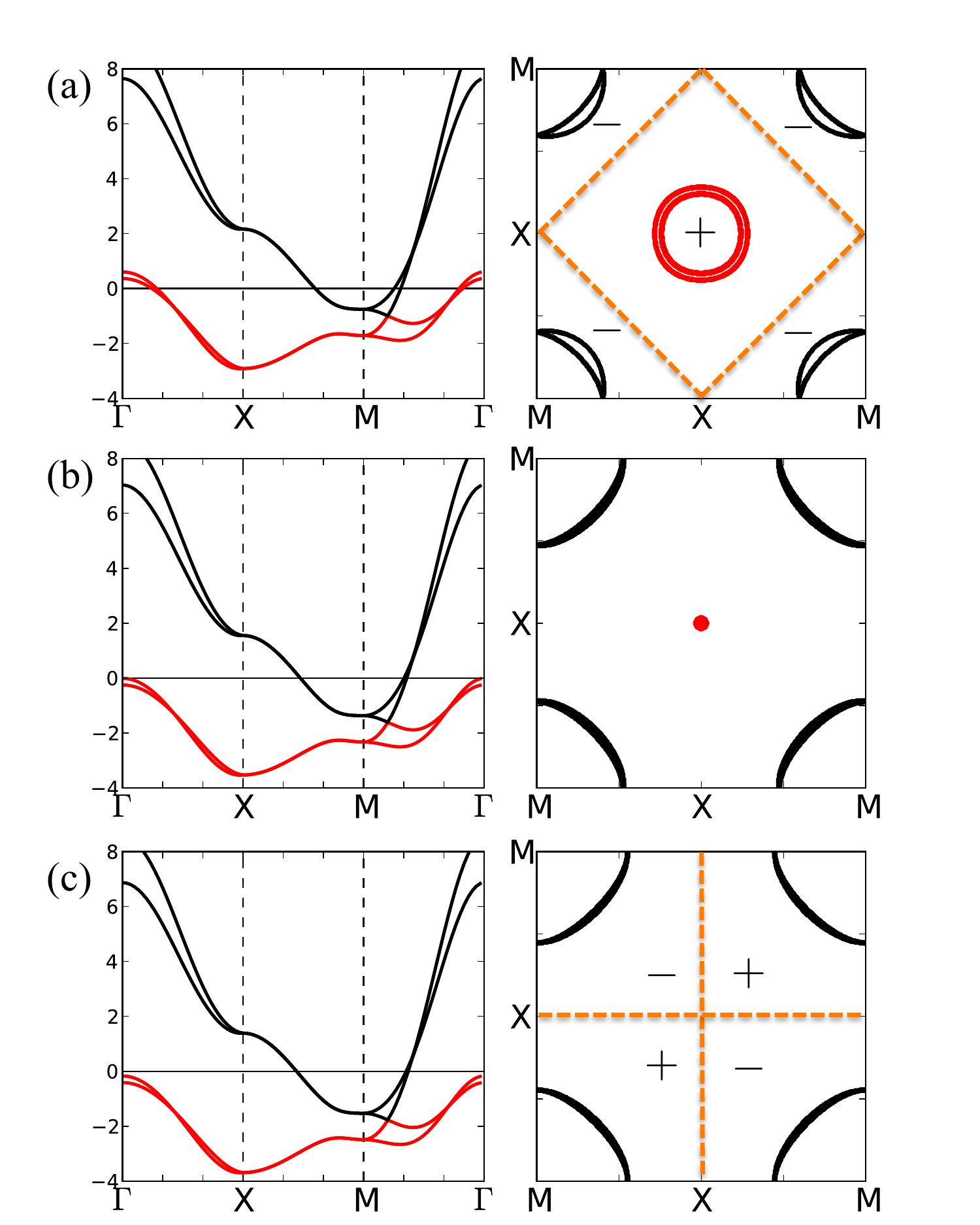}
  \caption
   {(Color online) The band structure and its corresponding Fermi surfaces at doping level $n=$ (a) $2.15$, (b) $2.4$, and (c) $2.46$ in the Brillouin zone of with two Fe atoms per unit cell. Black(red) lines represent the electron(hole) Fermi surface, and the orange (dashed) lines in (a)/(c) represent the nodal lines of s$_\pm$ / $d$-wave SC order $\Delta^\prime_{s_\pm}(k)$ / $\Delta_d(k)$ which are not crossing the Fermi surface. The plus and minus sign indicate the sign of the SC pairing order parameter in each region.}
   \label{fs}
\end{figure}

The FS topology is also responsible for the SC pairing symmetry of the system.
The zero-temperature FS is defined by zero energy contours of the quasiparticles, which can be drawn by using the Fourier transformation of the minimal hopping Hamiltonian.
In highly electron-doped samples, the SDW order is completely suppressed and thus we show the corresponding FSs in the BZ with two Fe atoms per unit cell.
In the following calculation we choose $V_{\text{NN}}=1.1$ and $V_{\text{NNN}}=1.05$, the same parameter as shown in the phase diagram Fig.~\ref{phase1}(a). Fig.~\ref{fs}(a) shows the FS in optimal-doped region, where $n=2.15$.
Here the hole pockets around $\Gamma$-point located at the center of the BZ can be clearly seen. This result is very similar to that in a previous work~\cite{28LPan}, which studied the evolution of the FS topology in FeAs-122 compound.
Fig.~\ref{fs}(b) shows the FS at $n=2.4$, where the $s_\pm$-wave SC is suppressed to zero and the hole pocket on the FS shrinks to a point at the center of the BZ.
We find that there is no hole pocket structure for all doping levels higher than $n=2.42$. Fig.~\ref{fs}(c) shows the FS at $n=2.46$, where k-space $d$-wave pairing order parameter $\Delta_{d}(k)=2\Delta_{d}\sin(k_x)\sin(k_y)$ is maximized.
The orange dashed lines here denote the nodes (or zeros) of $\Delta_{d}(k)$ and they do not cross the FS. In another word, $\Delta_{d}(k)$ is positive on the FS near $(\pi,\pi)$ and $(-\pi,-\pi)$, and negative near $(-\pi,\pi)$ and $(\pi,-\pi)$.
The sign change over neighboring electron pockets demonstrates that the SC phase from $n=2.4$ to $2.5$ shown in Fig.~\ref{phase1}(a) is of $d$-wave symmetry without nodes.
Our FS calculation result is in good agreement with a random phase approximation (RPA) calculation~\cite{30Das}.
It should be noticed that the pairing symmetry changes from $s_\pm$-wave to $d$-wave at the same doping level where the hole pocket disappears on the FS, which reveals the intrinsic correlation between FS structure and system's pairing symmetry.

We now proceed to study the effect of real-space inhomogeneity and address the outstanding question as to why the SC has not been observed in Ba(Fe$_{1-x}$Co$_x$As)$_2$ for $n>2.4$ where the SC has a $d$-wave symmetry according to the present calculation.
In this compound, there should be in addition to charge doping also significant scattering of the itinerant electrons due to randomly distributed Co atoms in the FeAs layer.
The disorder concentration can become rather densed when the sample is in the highly (Co) doped region.
We speculate that the densed disorder scattering suppress the $d$-wave SC in this compound.
On the other hand, the Fe-planes in A$_y$Fe$_2$Se$_2$ are quite clean because the doped A-atoms are between FeSe-layers, thus the impurity potential of the A atoms has little effect on electrons in the Fe-planes.
This is why the $d$-wave SC survives in highly electron-doped A$_y$Fe$_2$Se$_2$.
To consider the scattering effect of the disordered Co impurities in Ba(Fe$_{1-x}$Co$_x$As)$_2$, the Hamiltonian due to the impurity part can be written as
\begin{equation}
H_{imp}=V_{imp}\sum_{I\mu\sigma}c^\dagger_{I\mu\sigma}c_{I\mu\sigma},
\end{equation}
where $V_{imp}$ is the impurity strength at the $I$-th Co site in the lattice, the summation is over all randomly distributed impurity atoms.
In this work, the impurity potential of Co is known to be weaker than those of Ni or Cu~\cite{31Ideta}, and we set the impurity strength to be $V_{imp}=-1$.

\begin{figure}
 \includegraphics[scale=0.55,angle=0]{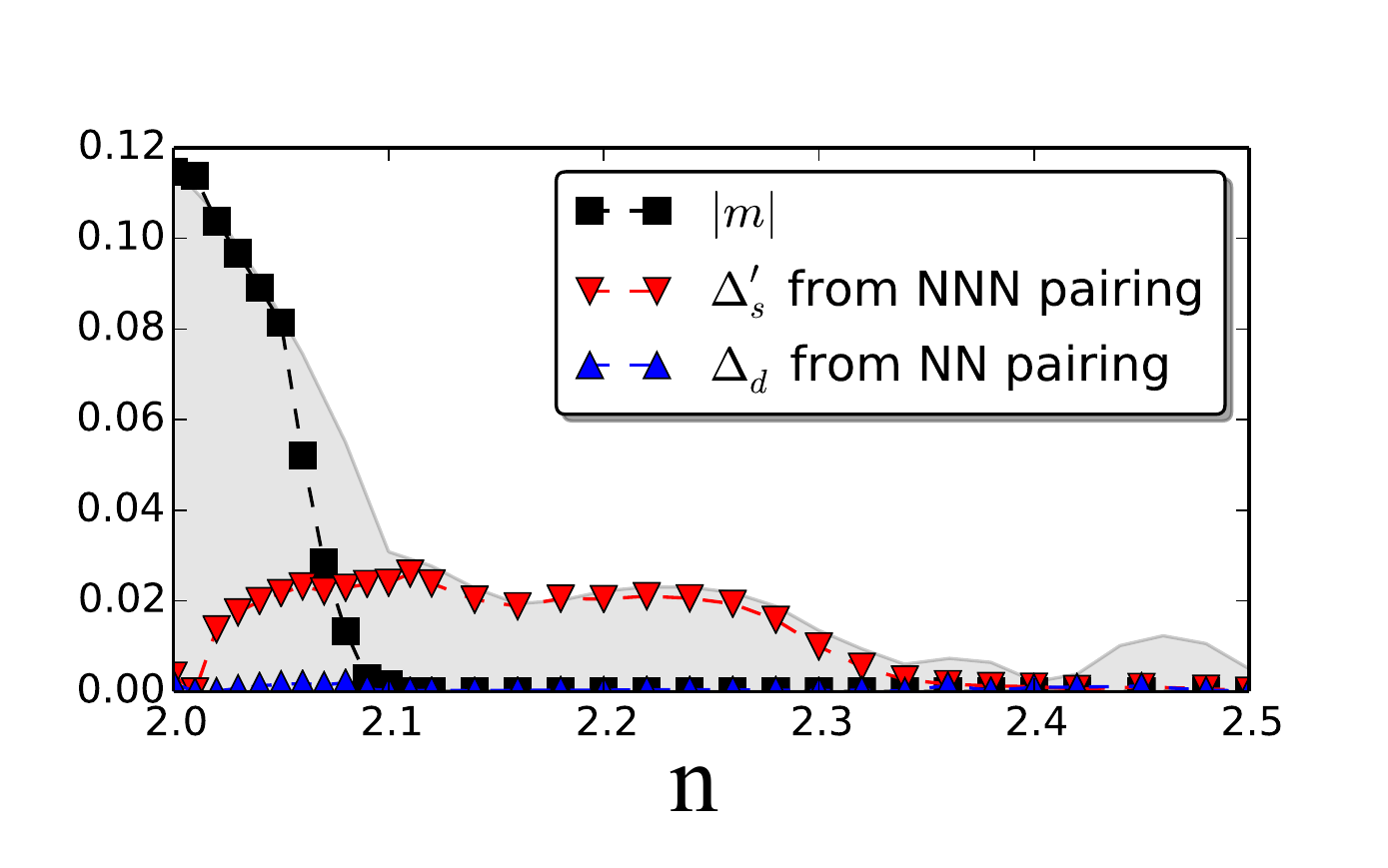}
  \caption
{(Color online) The phase diagram at $T=0K$ of Ba(Fe$_{1-x}$Co$_x$As)$_2$ as a function of $n$ after averaged over the randomly distributed Co impurities.
Black squares, red and blue triangles represent the collinear SDW order parameter, pairing order parameter of NNN $s_\pm$-wave and NN $d$-wave, correspondingly.
The impurity strength is set to be $V_{imp}=-1$, the NN and NNN pairings are set to be $V_{\text{NN}}=1.1$ and $V_{\text{NNN}}=1.05$.
The gray shaded background represents the boundary of the impurity-free calculated phase diagram which is the same as Fig.~\ref{phase1}(a).
}
   \label{phase2}
\end{figure}

Choosing the NN and NNN pairing to be $V_{\text{NN}}=1.1$ and $V_{\text{NNN}}=1.05$, the same parameter used to calculate the phase diagram in Fig.~\ref{phase1}(a), we have calculated the SDW and SC order parameters as a function of $n$ by averaging over $10$ different impurity configurations on a $28\times28$ lattice.
Our results for the phase diagram of Ba(Fe$_{1-x}$Co$_x$As)$_2$, after averaged over 10 impurity distribution configurations, are shown in Fig.~\ref{phase2} as a function of $n=2+x$ with $x$ as the concentration of doped electrons or Co impurities.
Since the statistical error bar at each point is smaller than the symbol itself, they are not shown on the graph.
Although the SC with $s_\pm$-wave pairing symmetry still exists in the region for $n<2.34$, the SC in the highly electron-doped region ($n>2.34$) is completely suppressed by the disordered Co atoms.
Here the $d$-wave SC exhibited in Fig.~\ref{phase1} for $n>2.4$ is destroyed and Andreev bound states are created by the impurities. Similar phase diagrams are also obtained for $V_{\text{NN}}=1.2$ and $1.3$, but we do not show them here.
The essential feature shown in Fig.~\ref{phase2} is consistent with experiments on Ba(Fe$_{1-x}$Co$_x$As)$_2$.
The predicted $d$-wave pairing symmetry for A$_y$Fe$_2$Se$_2$ with $0.8 \lesssim y \leq 1$ is very robust even when $V_{\text{NN}}>1.3$.

It is important to point out that the crystal structure of A$_y$Fe$_2$Se$_2$ with $y=1$ is identical to that of Ba(FeAs)$_2$.
With $1>y>0.8$, minor disorder is introduced into the A-layers.
But for $y=0.3$, the crystallographic structure may greatly deviate from that of Ba(FeAs)$_2$  and may not have a well-defined stable structure.
Since the FeSe layer is not affected by A-atom doping, the SC observed for $y=0.3$~\cite{27YTexier} should have s$_\pm$-pairing symmetry.
We predict that in A$_y$Fe$_2$Se$_2$ with $y=0.8$ to $1$, the pairing symmetry should be dominantly $d$-wave and the SC is of considerable strength if $V_{\text{NN}}>V_{\text{NNN}}$.

In this work, we presented a unified description of the evolution of superconductivity by including both the NN and the NNN intraorbital pairing interactions.
We showed that by starting with a phenomenological two-orbital model the pairing symmetry transforms from $s_\pm$-wave to $d$-wave  in highly electron-doped A$_y$Fe$_2$Se$_2$ as its Fermi surface topology changed.
The transition occurred when the hole pockets vanished near the $\Gamma$-point of the Brillouin zone as the hole states were completely filled by doped electrons.
We also found the emergence of a complex pairing state, $s_\pm+id$, in our calculations.
The existence of such a time-reversal symmetry breaking $s_\pm+id$ pairing state was suggested in previous studies from a pure band picture in k-space~\cite{CPlatt,MKhodas}.
However, we need to point out that our real-space formalism allows the study of the local spectra, as well as the stability and robustness of these competing phases in the presence of impurity scattering states, e.g., at very low impurity concentrations, or of vortex core states in magnetic fields.
We attribute the absence of $d$-wave superconductivity in highly electron-doped Ba(Fe$_{1-x}$Co$_x$As)$_2$ to the presence of randomly distributed Co impurities in the Fe square lattice with weak scattering potential.
In that sense Co doping is less destructive to superconductivity than Zn doping, which suppresses the s$_\pm$ pairing symmetry at around 10\% Zn concentration~\cite{HChen}. 

Besides the overall evolution of the superconducting symmetry from $s_\pm$-wave  to  $d$-wave pairing with doping, the other key result of our work is the detrimental effect of weak impurity scattering potential on superconductivity, namely at high Co concentrations.
Until now the Co-impurity scattering effect was ignored in calculations, mostly due to the nature of its weakness.
Here, we demonstrated that weak impurity scattering may play a crucial role for explaining the absence of $d$-wave pairing in some of the 122 iron-based superconductors at high electron doping.
Finally, we suggest experiments to probe the signature of the time-reversal symmetry breaking   $s_\pm+id$ state
in highly electron-doped samples, however, with impurity-free Fe planes. The most promising region in the phase diagram of our two-orbital model is therefore around electron filling of  $2.36\lesssim n <2.4$, where small hole pockets exist.

\acknowledgments

This work was supported in part by the Texas Center for Superconductivity at the University of Houston, the National Science Foundation through grant No. DMR-1206839 (B.L. \& K.E.B.), and the Robert A.Welch Foundation under Grant No. E-1146 (L.P. \& C.S.T.). Work at Los Alamos National Laboratory was supported by the U.S. DOE Contract No. DE-AC52-06NA25396 through the LANL LDRD Program (Y.-Y.T.) and  the U.S. DOE Office of Basic Energy Sciences (M.J.G. \& J.-X.Z.).

\end{document}